\documentclass[12pt]{article}
\usepackage{amscd,amsmath,amsxtra,amssymb,amstext,latexsym,amsthm}
\usepackage{graphicx}
\newcommand {\al}   {\alpha}       
\newcommand {\g }   {\gamma}       %\newcommand {\GM}  {\Gamma}
\newcommand {\dl}   {\delta}       \newcommand {\e }  {\epsilon}

\newcommand {\lm}   {\lambda}      
          
\newcommand {\s }   {\sigma}       
         
\newcommand {\vf }  {\varphi}      
         \newcommand {\om}  {\omega}

\newcommand {\pl}   {\partial}     \newcommand {\nb}  {\nabla}
\renewcommand {\sin}{{\sf\,sin\,}}       \renewcommand {\cos}{{\sf\,cos\,}}
\newcommand   {\ex}{{\sf\,e}}            
\newcommand   {\re}{{\sf\,re\,}}         
\newcommand   {\const}{{\sf\,const}}     
\newcommand {\MO}  {{\mathbb O}}   
   \newcommand {\MR}  {{\mathbb R}}
\newcommand {\MS}  {{\mathbb S}}   
   \newcommand {\Sd}  {{\textsc{d}}}

   \newcommand {\St}  {{\textsc{t}}}

\begin{document}
\title     {Torsional Elastic Waves in Cylindrical Waveguide with Wedge
            Dislocation}
\author    {M. O. Katanaev
            \thanks{E-mail: katanaev@mi.ras.ru}\\ \\
            \sl Steklov Mathematical Institute,\\
            \sl ul.~Gubkina, 8, Moscow, 119991, Russia}
\date      {25 February 2015}
\maketitle
\begin{abstract}
We describe propagation of torsional elastic waves in cylindrical waveguide
with wedge dislocation in the framework of geometric theory of defects. The
defect changes the dispersion relation. For positive deficit angles, it
increases the phase velocity, and decreases the group velocity of the wave.
\end{abstract}
%******************************************************************************
\section{Introduction}
%*******************************************************************************
Ideal crystals are absent in nature, and most of their physical properties, such
as plasticity, melting, growth, etc., are defined by defects of the crystalline
structure. Therefore, a study of defects is a topical scientific question of
importance for applications in the first place. At present, a fundamental theory
of defects is absent in spite of the existence of dozens of monographs
and thousands of articles.

One of the most promising approaches to the theory of defects is based on
Riemann--Cartan geometry, which involves nontrivial metric and torsion.
In this approach, a crystal is considered as a continuous elastic medium with
a spin structure. If the displacement vector field is a smooth function, then
there are only elastic stresses corresponding to diffeomorphisms of the
Euclidean space. If the displacement vector field has discontinuities, then
we are saying that there are defects in the elastic structure. Defects in the
elastic structure are called dislocations and lead to the appearance
of nontrivial geometry. Precisely, they correspond to a nonzero torsion tensor,
equal to the surface density of the Burgers vector. Defects in the spin
structure are called disclinations. They correspond to nonzero curvature tensor,
curvature tensor being the surface density of the Frank vector.

The idea to relate torsion to dislocations appeared in the 1950s [1--4].
\nocite{Kondo52,Nye53,BiBuSm55,Kroner58}
This approach is still being successfully developed (note reviews [5--11]),
\nocite{SedBer67,Kleman80A,Kroner81,DzyVol88,KadEde83,KunKun86,Kleine89}
and is often called the gauge theory of dislocations.

Some time ago we proposed the geometrical theory of defects [12--14]
\nocite{KatVol92,KatVol99,Katana05}. Our approach is essentially different from
others in two respects. Firstly, we do not have the displacement vector
field and rotational vector field as independent variables because, in general,
they are not continuous. Instead, the triad field and $\MS\MO(3)$-connection are
considered as independent variables. If defects are absent, then the triad and
$\MS\MO(3)$-connection reduce to partial derivatives of the displacement and
rotational angle vector fields. In this case the latter can be reconstructed.
Secondly, the set of equilibrium equations is different. We proposed purely
geometric set which coincides with that of Euclidean three dimensional gravity
with torsion. The nonlinear elasticity equations and principal chiral
$\MS\MO(3)$ model for the spin structure enter the model through the elastic and
Lorentz gauge conditions [14--16] \nocite{Katana03,Katana04,Katana05} which
allow to reconstruct the displacement and rotational angle vector fields
in the absence of dislocations in full agreement with classical models.

The advantage of the geometric theory of defects is that it allows one to
describe singele defects as well as their continuous distributions.

In the present paper, we consider propagation of torsional elastic waves in
cylindrical waveguide with wedge dislocation. It is a classical problem which
was solved in the absence of defect long ago within the elasticity theory
(see, for example, \cite{SneBer58}). We show that the wedge dislocation changes
the dispersion relation. For positive deficit angles the phase velocity
increases and group velocity decreases. Negative deficit angles have opposite
consequences.
%******************************************************************************
\section{Elastic waves in media with dislocations}
%*******************************************************************************
Let us introduce notation and remind necessary facts from differential geometry.
We consider elastic media as topologically trivial manifold which is
diffeomorphic to Euclidean space $\MR^3$. Denote Cartesian coordinate system
by $x^i$, $i=1,2,3$. Without dislocations, the elastic media is described by
the Euclidean metric $\dl_{ij}$. If dislocations are present then the metric
becomes nontrivial:
\begin{equation*}
  \dl_{ij}\mapsto g_{\mu\nu}(x)=e_\mu{}^ie_\nu{}^j\dl_{ij},\qquad\mu,\nu=1,2,3,
\end{equation*}
where $e_\mu{}^i(x)$ is the triad. The triad field defines the orthonormal basis
of the tangent spaces at each point $e_i:=e^\mu{}_i\pl_\mu$, where $e^\mu{}_i$
is the inverse triad. We shall see that this basis is more convenient in
applications.

Assume that relative displacements for elastic deformations are much smaller
then deformations caused by defects:
\begin{equation}                                                  \label{edueco}
  \pl_\mu u^i\ll e_\mu{}^i.
\end{equation}
where $u^i(t,x)$ are components of the displacement vector field with respect to
orthonormal basis in the tangent space $e_i$. Then in the first approximation,
the elastic waves propagate in Riemannian space with nontrivial metric produced
by dislocations. Here we neglect changes in the metric produced by elastic waves
themselves. Therefore for elastic waves in media with defects we postulate the
following wave equation
\begin{equation}                                                  \label{ephops}
  \rho_0\ddot u^i-\mu\tilde\triangle u^i-(\lm+\mu)
  \widetilde\nb^i\widetilde\nb_ju^j=0,
\end{equation}
where $\lm$ and $\mu$ are Lame coefficients,
$\tilde\triangle:=\widetilde\nb^i\widetilde\nb_i$ is the covariant
Laplace--Beltrami operator for the triad field $e_\mu{}^i$, and
$\widetilde\nb_i$ is the covariant derivative. The explicit form of the
covariant derivative is
\begin{equation}                                                  \label{qmjfuy}
  \widetilde\nb_iu^j:=e^\mu{}_i\widetilde\nb_\mu u^j
  =e^\mu{}_i(\pl_\mu u^j+u^k\widetilde\om_{\mu k}{}^j),
\end{equation}
where $\widetilde\om_{\mu k}{}^j$ is the $\MS\MO(3)$-connection constructed for
zero torsion.

Note that the displacement vector field $u^i$ is not the total displacement
vector field of points in the media with dislocations. In those regions where
defects are absent, the total displacement vector is $u^i_\Sd+u^i$ where
$u^i_\Sd$ is the dislocation part of the displacement vecor field which produces
the triad $e_\mu{}^i:=\pl_\mu u^i_\Sd$. Note also that the smallness of relative
elastic deformations (\ref{edueco}) is meaningful at those regions where
displacements $u^i_\Sd$  are not defined (they are not continuous functions if
dislocations are present).

The expression for the deformation tensor $\s_{ij}$ is needed to pose the
boundary value problem for wave equation (\ref{ephops}). In the presence of
defects the deformation tensor is defined as follows
\begin{equation}                                                  \label{qnbhyx}
  \e_{ij}:=\frac12(e^\mu{}_i\widetilde\nb_\mu u_j
  +e^\mu{}_j\widetilde\nb_\mu u_i).
\end{equation}

Thus the existence of dislocations in media results in the introduction of the
triad field $e_\mu{}^i(x)$, metric $g_{\mu\nu}(x)$, and replacement of partial
derivatives by covariant ones.

Remember that lowering of Latin indices is performed using the Euclidean metric,
$u_i:=\dl_{ij}u^j$, and it commutes with covariant differentiation.
%******************************************************************************
\subsection{Torsional waves in cylindrical waveguide with wedge dislocation}
\begin{figure}[h,b,t]%----------------------------------------------------------
\hfill\includegraphics[width=.4\textwidth]{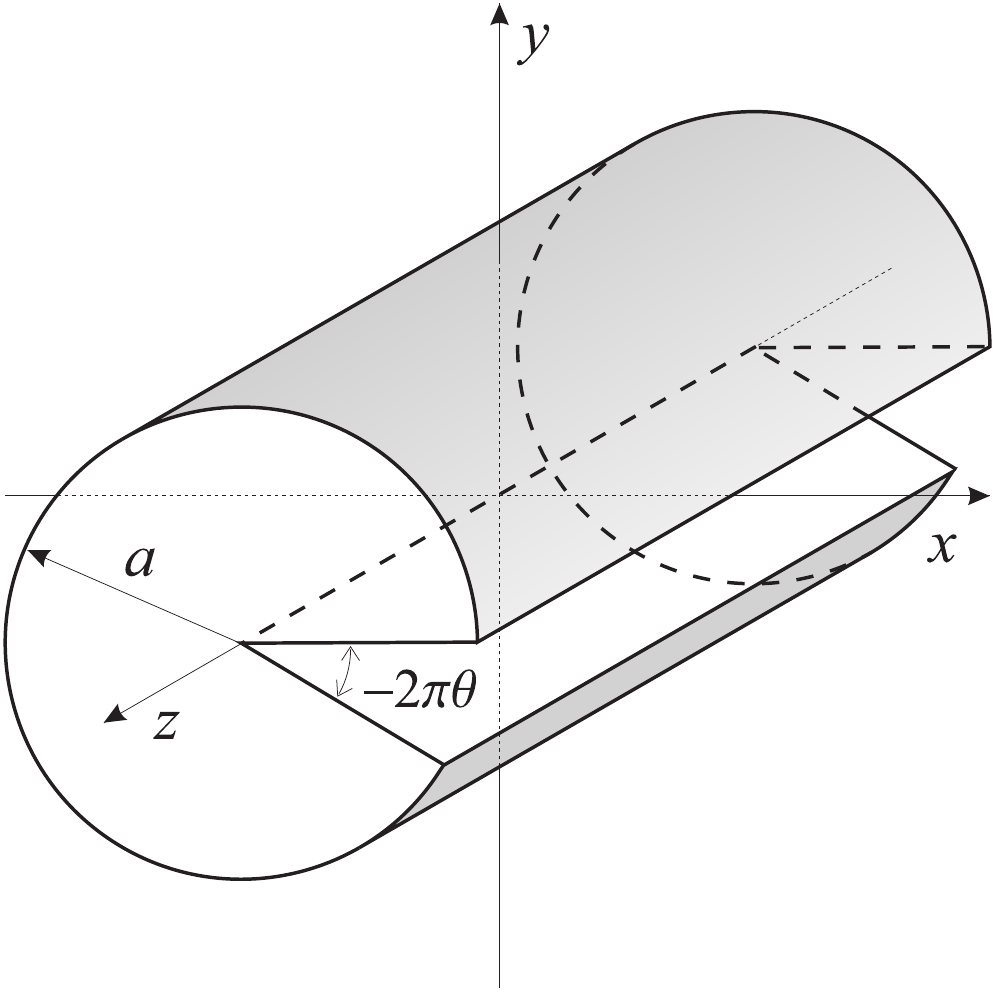}
\hfill {}
\centering\caption{Wedge dislocation with the deficit angle $2\pi\theta$. For
 negative and positive $\theta$, the wedge is cut out or added, respectively.}
\label{fwedis}
\end{figure}

%*******************************************************************************
We consider cylindrical waveguide of radius $a$ with a wedge dislocation. Its
axis is assumed to coincide with the axis of cylinder. We choose cylindrical
coordinate system $r,\vf,z$, the $z$ axis coinciding with the cylinder axis as
well. Let the wedge dislocation has deficit angle $2\pi\theta$ (see
Fig.\ref{fwedis})
%-------------------------------------------------------------------
If $-1<\theta<0$, then the wedge is cut out from the cylinder. Dislocation is
absent for $\theta=0$. For positive deficit angle, $\theta>0$, the wedge of
media is added to the cylinder.

The cylinder contains wedge dislocation, and it creates nontrivial metrics. The
metric becomes noneuclidean \cite{Katana03}
\begin{equation}                                                  \label{qnhsth}
  ds^2=\left(\frac ra\right)^{2\g-2}\left(dr^2+\frac{\al^2r^2}{\g^2}d\vf^2
  \right)+dz^2,
\end{equation}
where $\al:=1+\theta$, and we introduce dimensionless constant
\begin{equation}                                                  \label{qmnjfy}
  \g:=-\theta b+\sqrt{\theta^2b^2+1+\theta}.
\end{equation}
The constant
\begin{equation*}
  b:=\frac\s{2(1-\s)}
\end{equation*}
is defined by dimensionless Poisson ratio
\begin{equation*}
  \s:=\frac\lm{2(\lm+\mu)},
\end{equation*}
which characterize elastic properties of media. Thermodynamical arguments
restrict possible values of Poisson ratio to $-1\le\s\le1/2$ \cite{LanLif70}.

The limit
\begin{equation}                                                  \label{qnbhyi}
  \theta\to0,\qquad\al\to1,\qquad\g\to1
\end{equation}
corresponds to absence of the dislocation.

Metric (\ref{qnhsth}) is rotationally and translationally invariant.
Two dimensional metric on sections $z=\const$ is the metric of conical
singularity which arises due to the wedge dislocation. It is important that this
metric is written in elastic gauge and therefore is consistent with the
elasticity theory \cite{Katana03}.

Metric for the wedge dislocation (\ref{qnhsth}) is defined for all deficit
angles $\theta>-1$ and all values of radial coordinate $0<r<a$. It is degenerate
on the dislocation axis $r=0$. Metric (\ref{qnhsth}) coincide with the induced
metric in classical elasticity theory only for small deficit angles and near the
boundary of the cylinder $r\sim a$ \cite{Kosevi81,Katana03}.

Metric for the wedge dislocation defines the triad field which we choose to be
diagonal:
\begin{equation}                                                  \label{qbndhy}
  e_r{}^{\hat r}=\left(\frac ra\right)^{\g-1},\qquad
  e_\vf{}^{\hat\vf}=\left(\frac ra\right)^{\g-1}\frac{\al r}\g,\qquad
  e_z{}^{\hat z}=1,
\end{equation}
where $\mu=r,\vf,z$ and $i=\hat r,\hat\vf,\hat z$.
In the absence of dislocation, $\g=1,\al=1$, it defines the usual orthonormal
basis for tangent spaces in cylindrical coordinates. The inverse triad is
\begin{equation}                                                  \label{qbvxgt}
  e^r{}_{\hat r}=\left(\frac ar\right)^{\g-1},\qquad
  e^\vf{}_{\hat\vf}=\left(\frac ar\right)^{\g-1}\frac \g{\al r},\qquad
  e^z{}_{\hat z}=1.
\end{equation}

To find the explicit form of the wave operator, we need to compute Christoffel's
symbols and components of $\MS\MO(3)$-connection. Straightforward calculations
show that only for Christoffel's symbols differ from zero:
\begin{equation}                                                  \label{qbnxgs}
\begin{split}
  \widetilde\Gamma_{rr}{}^r&=\frac{\g-1}r,
\\
  \widetilde\Gamma_{r\vf}{}^\vf=\widetilde\Gamma_{\vf r}{}^\vf&=\frac\g r,
\\
  \widetilde\Gamma_{\vf\vf}{}^r&=-\frac{\al^2 r}\g.
\end{split}
\end{equation}
Triad (\ref{qbndhy}) defines also components of $\MS\MO(3)$-connection
\begin{equation*}
  \om_{\mu i}{}^j=-\pl_\mu e_\nu{}^j e^\nu{}_i
  +e^\nu{}_i\widetilde\Gamma_{\mu\nu}{}^\rho e_\rho{}^j.
\end{equation*}
Only two components differ from zero:
\begin{equation}                                                  \label{qmnjpo}
  \om_{\vf\hat r}{}^{\hat\vf}=-\om_{\vf\hat\vf}{}^{\hat r}=\al.
\end{equation}

Now we can compute the Laplacian
\begin{multline}                                                  \label{qbnxht}
  \tilde\triangle u_i=g^{\mu\nu}\pl^2_{\mu\nu} u_i
  -g^{\mu\nu}\pl_\mu\om_{\nu i}{}^j u_j-2g^{\mu\nu}\om_{\mu i}{}^i\pl_\nu u_j-
\\
  -g^{\mu\nu}\widetilde\Gamma_{\mu\nu}{}^\rho(\pl_\rho u_i-\om_{\rho i}{}^ju_j)
  +g^{\mu\nu}\om_{\mu i}{}^k\om_{\nu k}{}^ju_j.
\end{multline}
Substitution of explicit expressions for Christoffel's symbols (\ref{qbnxgs})
and $\MS\MO(3)$-connection (\ref{qmnjpo}) yields the following expressions:
\begin{equation}                                                  \label{qmkhsq}
\begin{split}
  \tilde\triangle u_{\hat r}&=\left(\frac ar\right)^{2\g-2}\frac1r\pl_r
  \big(r\pl_r u_{\hat r}\big)+\left(\frac ar\right)^{2\g-2}\frac{\g^2}{\al^2r^2}
  \pl^2_{\vf\vf}u_{\hat r}+\pl^2_{zz}u_{\hat r}-
\\
  &\qquad\qquad\qquad\qquad
  -\left(\frac ar\right)^{2\g-2}\frac{\g^2}{r^2}u_{\hat r}
  -2\left(\frac ar\right)^{2\g-2}\frac{\g^2}{\al r^2}\pl_\vf u_{\hat\vf},
\\[2mm]
  \tilde\triangle u_{\hat\vf}&=\left(\frac ar\right)^{2\g-2}\frac1r\pl_r
  \big(r\pl_ru_{\hat\vf}\big)+\left(\frac ar\right)^{2\g-2}\frac{\g^2}{\al^2 r^2}
  \pl^2_{\vf\vf}u_{\hat\vf}+\pl^2_{zz}u_{\hat\vf}-
\\
  &\qquad\qquad\qquad\qquad
  -\left(\frac ar\right)^{2\g-2}\frac{\g^2}{r^2}u_{\hat\vf}
  +2\left(\frac ar\right)^{2\g-2}\frac{\g^2}{\al r^2}\pl_\vf u_{\hat r},
\\[2mm]
  \tilde\triangle u_{\hat z}&=\left(\frac ar\right)^{2\g-2}\frac1r\pl_r
  \big(r\pl_r u_{\hat z}\big)+\left(\frac ar\right)^{2\g-2}\frac{\g^2}{\al^2r^2}
  \pl^2_{\vf\vf}u_{\hat z}+\pl^2_{zz}u_{\hat z}.
\end{split}
\end{equation}

Corresponding wave equations (\ref{ephops}) have many solutions. The simplest
one describes torsional waves for which only the angular component $u_{\hat\vf}$
differs from zero and does not depend on the angular coordinate $\vf$:
\begin{equation*}
  u_{\hat r}=0,\qquad u_{\hat\vf}=u_{\hat\vf}(t,r,z),\qquad u_{\hat z}=0.
\end{equation*}
It is easy to see that torsional waves take place without media compression
\begin{equation*}
  \e:=\widetilde\nabla_iu^i=0.
\end{equation*}
For these oscillations, wave equations $\widetilde\square u_{\hat r}=0$ and
$\widetilde\square u_{\hat z}=0$ are identically satisfied, and we are left with
one wave equation
\begin{equation}                                                  \label{qbnckj}
  \frac1{c_\St^2}\ddot u_{\hat\vf}-\left(\frac ar\right)^{2\g-2}\frac1r\pl_r
  \big(r\pl_r u_{\hat\vf}\big)-\pl^2_{zz}u_{\hat\vf}
  +\left(\frac ar\right)^{2\g-2}\frac{\g^2}{r^2}u_{\hat\vf}=0.
\end{equation}

We look for solution of this equation in the plain wave form
\begin{equation*}
  u_{\hat\vf}=\re\left(U\ex^{i(kz-\om t)}\right),
\end{equation*}
where $U(r)\in\MR$ is the wave amplitude, $k\in\MR$ is the wave vector, and
$\om\in\MR$ is the wave frequency. Then wave equation (\ref{qbnckj}) takes the
form
\begin{equation}                                                  \label{qnjirn}
  r\pl_r\left(r\pl_r U\right)+\kappa^2a^2\left(\frac ra\right)^{2\g}U-
  \g^2 U=0,
\end{equation}
where
\begin{equation}                                                  \label{qnhdyt}
  \kappa^2:=\frac{\om^2}{c_\St^2}-k^2.
\end{equation}
Now we introduce new radial coordinate
\begin{equation*}
  r=ar'^{\frac1\g},\qquad 0<r'<1.
\end{equation*}
Then equation (\ref{qnhdyt}) reduces to the Bessel equation
\begin{equation}                                                  \label{qnnmcj}
  r'^2\frac{d^2 U}{dr'^2}+r'\frac{dU}{dr'}+\lm^2r'^2U-U=0,
\end{equation}
where
\begin{equation*}
  \lm^2:=\frac{\kappa^2a^2}{\g^2}.
\end{equation*}
A general solution to the Bessel equation has two integration constants.
We require it to be finite at $r=0$. Then the amplitude takes the form
\begin{equation}                                                  \label{qzaqws}
  U=AJ_1\left(\lm\left(\frac ra\right)^\g\right),
\end{equation}
where $A\in\MR$ is the integration constant and $J_1$ is the Bessel function of
the first kind of order one (see, for example, \cite{JaEmLo60}).

If dislocation is absent, $\g=1$, then the solution takes the well known form
(see, for example, \cite{SneBer58})
\begin{equation*}
  U\underset{\g\to1}\to AJ_1(\kappa r)
\end{equation*}

To impose boundary conditions on the cylinder surface, we need explicit form
of deformation tensor (\ref{qnbhyx}). Straightforward calculations yeild the
following expressions:
\begin{equation}                                                  \label{qmjifj}
\begin{aligned}
  \e_{\hat r\hat r}&=\left(\frac ar\right)^{\g-1}\pl_r u_{\hat r}, & \qquad
  \e_{\hat r\hat\vf}&=\frac12\left(\frac ar\right)^{\g-1}\left[\pl_ru_{\hat\vf}
  +\frac\g{\al r}\pl_\vf u_{\hat r}-\frac \g ru_{\hat\vf}\right],
\\
  \e_{\hat\vf\hat\vf}&=\left(\frac ar\right)^{\g-1}\frac\g{\al r}
  \big(\pl_\vf u_{\hat\vf}+\al u_{\hat r}\big), &
  \e_{\hat r\hat z}&=\frac12\left[\left(\frac ar\right)^{\g-1}\pl_r u_{\hat z}
  +\pl_z u_{\hat r}\right],
\\
  \e_{\hat z\hat z}&=\pl_z u_{\hat z}, &
  \e_{\hat\vf\hat z}&=\frac12\left[\left(\frac ar\right)^{\g-1}\frac\g{\al r}
  \pl_\vf u_{\hat z}+\pl_z u_{\hat\vf}\right].
\end{aligned}
\end{equation}
Only two components
\begin{equation*}
  \e_{\hat r\hat\vf}=\e_{\hat\vf\hat r}=\frac12\left(\frac ar\right)^{\g-1}
  \left[\pl_ru_{\hat\vf}-\frac \g ru_{\hat\vf}\right]
\end{equation*}
differ from zero for torsional waves. We require the cylinder surface to be
free. Then the elastic forces on the boundary must be zero. It yeilds the
boundary condition
\begin{equation}                                                  \label{qmnjia}
  \left[\pl_ru_{\hat\vf}-\frac \g ru_{\hat\vf}\right]_{r=a}.
\end{equation}
For solution (\ref{qzaqws}), it takes the form
\begin{equation*}
  \lm J'_1(\lm)-J_1(\lm)=0,
\end{equation*}
where prime denotes differentiation of the Bessel function with respect to its
argument. There is equality
\begin{equation*}
  J'_1(\lm)=J_0(\lm)-\frac1\lm J_1(\lm).
\end{equation*}
Then the boundary condition is equivalent to the equality which defines the
dispersion relation
\begin{equation}                                                  \label{qbnzna}
  \frac{\kappa a}\g=\xi \qquad\Leftrightarrow\qquad
  \om=c_\St\sqrt{k^2+\frac{\g^2\xi^2}{a^2}}.
\end{equation}
where $\xi$ is a root of the equation
\begin{equation}                                                  \label{qbnlna}
  \xi J_0(\xi)=2J_1(\xi).
\end{equation}

Bessel functions $J_\nu$ have the following asymptopics for large argument,
$\xi\gg1$, $\xi\gg\nu$,
\begin{align*}
  J_0(\lm)&\approx\sqrt{\frac2{\pi\lm}}\cos\left(\lm-\frac\pi4\right),
\\
  J_1(\lm)&\approx\sqrt{\frac2{\pl\lm}}\sin\left(\lm-\frac\pi4\right).
\end{align*}
Then equation (\ref{qbnzna}) takes the form
\begin{equation*}
  \xi\cos\left(\xi-\frac\pi4\right)=2\sin\left(\xi-\frac\pi4\right)
\end{equation*}
for large argument. This equation has countable set of roots, each of them
defining the dispersion relation.

The phase velocity of torsional waves $v:=\om/k$ has the form
\begin{equation}                                                  \label{qbmnzs}
  v=c_\St\sqrt{1+\frac{\g^2\xi^2}{a^2 k^2}}.
\end{equation}
It is not difficult to calculate the group velocity
\begin{equation}                                                  \label{qbvgyt}
  v_{\rm g}:=\frac{d\om}{dk}=\frac{c^2_\St}v.
\end{equation}

If dislocation is absent, equation (\ref{qbnzna}) for  $\xi$ remains the
same. Relation  (\ref{qbvgyt}) between phase and group velocities preserves its
form. Thus the presence of the wedge dislocation changes only dispersion
relation (\ref{qbnzna}).

For positive deficit angles $\g>1$, the phase velocity increases and the group
velocity decreases as the consequence of equations (\ref{qbmnzs}) and
(\ref{qbvgyt}).

For small deficit angles the first correction has the form
\begin{equation*}
  \g\approx1+\theta\frac{1-2\s}{2(1-\s)}
\end{equation*}
and
\begin{equation}                                                  \label{qmikyh}
  \om\approx c_\St\sqrt{k^2+\frac{\xi^2}{a^2}}\left[1+\theta\frac{1-2\s}{2(1-\s)}
  \frac1{1+\frac{k^2a^2}{\xi^2}}\right].
\end{equation}
It is linear in deficit angle of the dislocation.
%******************************************************************************
\section{Conclusion}
%*******************************************************************************
We showed that the presence of the wedge dislocation in cylindrical waveguide
leads to changing of the dispersion relation for torsional elastic waves. The
difference increases as long as the deficit angle increases. For positive
deficit angle, the phase velocity of the wave increases and group velocity
decreases. For negative deficit angle, the behavior is opposite.

Solution of this
problem within the classical elasticity encounters essential technical problems.
Indeed, relative displacements are not small in the neighbourhood of the
dislocation axis even for small deficit angles, and the expression for the
induced metric is much more complicated \cite{Kosevi81,Katana03}. Solution of
this problem within the geometric theory of defects as simpler.

This work is supported by the Russian Science Foundation (project 14-11-00687)
in Steklov Mathematical Institute.
%\end{document}

\end{document}